\documentstyle[aps,preprint]{revtex}

\author{Diego F. Torres\thanks{e-mail: dtorres@venus.fisica.unlp.edu.ar} and H\'ector Vucetich}
\address{Departamento de F\'{\i}sica, Universidad Nacional de La Plata,
C.C. 67, 1900, La Plata, Buenos Aires, Argentina}
\title{Hyperextended Scalar-Tensor Gravity}

\begin{document}

\maketitle
\begin{abstract}
We study a general Scalar-Tensor Theory
with an arbitrary coupling funtion $\omega
(\phi )$ but also an arbitrary dependence of the {\sl gravitational constant}
$G(\phi )$ in the cases in which either one of them, or both, do not
admit an analytical inverse,
as in the hyperextended inflationary scenario.
We present the full set of field equations and
study their cosmological behavior.
We show that different scalar-tensor theories can be grouped in classes
with the same solution for the scalar field.

PACS {\it number(s)}: 04.50.+h, 04.20.Cv, 98.80.Cq, 98.80.Hw

\end{abstract}

\newpage

\section{Introduction}

Scalar-Tensor theories of gravity have an interesting physical embodiment
which makes them a natural generalization of General Relativity (GR). This
provide a convenient framework for the study of observational limits on
possible deviation of Einstein's theory, making them a profitable arena for
cosmology.

The archetypical and best known case of Scalar-Tensor Theory is Brans-Dicke
Gravity (BD) \cite{BransDicke} where there is a coupling function $\omega
(\phi )$ equal to a constant. More general cases with more complicated
couplings have also been studied \cite{STG}. In any case, in order to
evaluate the cosmological scenario and to test the predictable force of any
Scalar-Tensor Theory, it is necessary to have exact analytical solutions of
the field equations. Once having these solutions, simultaneous constraints
arising from different epochs of cosmic history must been set up. That is
the case for primordial nucleosynthesis \cite{Nucleo} and the weak-field
solar system test \cite{Will}. It has also been shown that Scalar-Tensor
theories may drive new forms of inflation \cite{Ext,Hyperext} and that
unusual physical effects arise on black hole physics if the {\em gravitational
constant} becomes a scalar field dependent magnitude \cite{BHole}. On the
other hand, perhaps a more philosophical way of thinking about Scalar-Tensor
theories of gravity is related to the Mach's Principle and the nature of
space and the inertial properties of the bodies.
Comparatively, little advance have been reached in this area up to
date \cite{MAch}. Scalar-Tensor theories have also been related with
strings, in which a dilaton field coupled to the curvature appears in the
low energy effective action \cite{Fradkin}.

Recently, a great improvement in the search of solutions of the field
equations have been given in the form of methods that allow analytical
integration through suitable changes of variables. Barrow \cite{Barrow1}
presented a method which enables exact solutions to be found for vacuum and
radiation dominated Friedmann universes of all curvatures in arbitrary
coupling Scalar-Tensor theories. Then, and also for arbitrary $\omega (\phi
) $, Barrow and Mimoso \cite{Barrow2} and Mimoso and Wands \cite{MW} derived
exact Friedmann-Robertson-Walker (FRW) cosmological solutions in models with
a perfect fluid satisfying the equation of state $p=(\gamma -1)\ \rho $
(with $\gamma $ a constant and $0\leq \gamma \leq 2$).

However, Scalar-Tensor theories have been formulated in two different ways
depending on the choice of the basic action or, equivalently, of the
lagrangian density for the field.
Via a field redefinition one can
establish the equivalence between these lagrangians (see below) and so
between the theories of gravitation they lead. But, as was clearly remarked
by Liddle and Wands \cite{LiddleWands} this is
not always possible. So, we have two physically different theories arising
from the fact that, in the general case, we have two non-related functions
of the field $\phi $; {\it i.e.} $G(\phi )$ and the coupling $\omega (\phi )$%
; where $G(\phi )$ is not limited to the form $1/\phi $ but
it is an arbitrary function of the field.
Since there is a deep connection between these models and Hyperextended
Inflation we propose to call Hyperextended Scalar-Tensor gravity to
these kind of two free functions theories.

In this work, we study the equivalence among the different scalar
lagrangians densities coupled to gravity that may be constructed
retaining only a term proportional to the curvature scalar. We present
the field equations for the more general Scalar-Tensor Theory, {\it
i.e. }with arbitrary dependence of $\omega (\phi )$, $G(\phi )$ and
eventually a potential term $V(\phi )$ and show how to extend the
procedure described in \cite{MW} to analytically solve the system of
the field equations in any of the geometries of space time. As in
\cite{Barrow1},\cite{Barrow2} and \cite {MW} the solutions will be
given in terms of a single integral over $\phi $ which may be
performed exactly in many cases (namely, in the cases of vacuum,
radiation and stiff filled universes) and numerically in all cases.

The paper is organized as follows. In Section II we describe the equivalence
problems among lagrangians; Section III presents the field equations and
in Section IV the FRW models 
are introduced together with a convenient choice of
variables. The procedure to obtain cosmological solutions is shown in V. 
Finally,  our conclusions are
sketched in VI.

\section{Equivalence among scalar lagrangians densities}

The more general lagrangian density for a scalar field coupled to gravity in
the usual way; {\it i.e. }it has only a term proportional to the curvature
scalar, is:

\begin{equation}
\label{1}L=16\pi L_M+\frac{K(\phi )}2\phi _{,\mu }\ \phi ^{,\mu }+G(\phi
)^{-1}R+V(\phi )
\end{equation}
where $L_M$ represents the lagrangian density for the matter content of the
space-time with no dependence on $\phi $ and $K, G^{-1}=1/G$ and $V$ are
arbitrary functions of the field.

In general, it is common to find in the literature only one of the two
following lagrangians densities:

\begin{equation}
\label{2}L_1=16\pi L_M+\phi \ R-\frac{\omega (\phi )}\phi \phi _{,\mu }\
\phi ^{,\mu }-V(\phi )
\end{equation}

\begin{equation}
\label{3}L_2=16\pi L_M+f(\phi )\ R+\frac 12\phi _{,\mu }\ \phi ^{,\mu
}-V(\phi )
\end{equation}
$L_1$ leads to the well known generalized Brans-Dicke theories of gravity
\cite{STG} while $L_2$ is referred to as a non-minimally coupled gravity. 
As particular cases of $L$ we shall have $L_1$ reproduced
when $K(\phi )=-2\,\omega (\phi )\,/\phi $ and $G(\phi )=1\,/\phi $
simultaneously (the BD cases) and $L_2$ when $K(\phi )=1$.
$L_{1\text{ }}$and $L_2$ are related through a scalar field transformation
which may be completed defining another field $\psi $ by:

\begin{equation}
\label{4}\psi =f(\phi )
\end{equation}
Defining also the coupling as:

\begin{equation}
\label{5}\omega =-\frac 12\ \frac{f(\phi )}{\left( \frac{df}{d\phi }\right) ^2%
}\,
\end{equation}
$L_2$ may be transformed to the form of $L_1$ for the new field $\psi $.
This kind of transformation was first noted by Nordtvedt \cite{STG} and
usually recalled by almost all the workers in the area. In particular,
Steinhardt and Ascetta \cite{Hyperext} used this transformation to study the
mechanism of Hyperextended Inflation. However, it is easy to see that we
have here a dependence on the {\it simplicity} of the coupling or the
functional form of $G(\phi )$. As it is noted in \cite{LiddleWands} if one
takes $\omega (\phi )=\omega _0+\omega _m\phi ^m$ (as in \cite{Barrow3}) or $%
f(\phi )$ as a truncated Taylor series (as in \cite{Hyperext}) one cannot
write down the equivalence between $L_1$ and $L_2$; in fact, to do such a
thing one has to ask for the existence of the analytical inverse of $f(\phi
) $ (note that $f\equiv 1/G$). So, the
choice of Steinhardt and Ascetta leads to a singularity in the $\phi -\psi $
transformation and this constitute the representation of a physical
difference between the two lagrangians densities. In these cases, and in
general, in all cases in which $G(\phi )$ is not an analytically invertible
function of $\phi $ the basic actions differs and so the theory of gravity
they lead and the cosmological effects of it.
A similar situation comes down when one tries to
establish an equivalence between $L$ and $L_1$ - $L_2$.

\section{Hyperextended Scalar-Tensor Theories}

>From now on, we shall call $K(\phi )=\frac{-2\omega (\phi )}\phi $ to
facilitate comparison with the BD cases by only particularizing the
dependence of $G(\phi )$. Anyway, this does not
represents a loss of generality but
only a change in the names of the functions. Taking variational derivatives
of the action constructed using the lagrangian density (\ref{1})
with respect to the dynamical variables $g^{\mu \nu }$ and $\phi $
yields to the field equations \cite{Diploma}:

\begin{equation}
\label{8}R_{\mu \nu }-\frac 12g_{\mu \nu }R=G(\phi )\left[ 8\pi T_{\mu \nu
}+\frac \omega \phi \phi _{,\mu }\,\phi ^{,\mu }-\frac \omega {2\phi }\phi
_{,\alpha }\,\phi ^{,\alpha }g_{\mu \nu }-\frac{V}{2}\,g_{\mu \nu }+
(G^{-1})_{,\mu;\nu }-g_{\mu \nu }\Box (G^{-1})\right]
\end{equation}
\begin{equation}
\label{9}R\ \frac{dG^{-1}}{d\phi }+\frac 1\phi \frac{d\omega }{d\phi }\phi
_{,\mu }\,\phi ^{,\mu }-\frac \omega {\phi ^2}\phi _{,\mu }\,\phi ^{,\mu }+
\frac{2\omega }\phi \ \Box \phi -\frac{dV}{d\phi }=0
\end{equation}
The second equation may be written down in a more usual way 
which involves the trace of
the stress-energy tensor of matter fields instead of the curvature
scalar. 

It is very important to remark that the usual relation $T_{\quad ;\nu }^{\mu
\nu }=0$ establishing the conservation laws (in the meaning of GR) of the
matter fields holds true. This may be seen by direct differentiation from
(\ref{8}) recalling the identities of the curvature tensor as a commutator of
covariant derivatives.

\section{Friedmann-Robertson-Walker Models}

We shall consider homogeneous and isotropic models with the metric given
by the Friedmann-Robertson-Walker (FRW) line element:

\begin{equation}
\label{11}ds^2=dt^2-a(t)^2\left[ \frac{dr^2}{1-kr^2}+r^2\left( d\theta
^2+\sin {}^2\theta \ d\Phi ^2\right) \right]
\end{equation}
In this framework, all the scalars are functions only of time
and not of the space coordinates. As equation of state we shall
use that of a perfect fluid $p=(\gamma -1)\ \rho $ (with $\gamma $ a
constant and $0\leq \gamma \leq 2$). The field equations become:

\begin{equation}
\label{12}\left( \frac{\dot a}a\right) ^2-\left( \frac{\dot a}a\right) \frac
1G\frac{dG}{d\phi }\dot \phi -\frac \omega 6\frac{\dot \phi ^2}\phi G+\frac
k{a^2}=\frac{8\pi }3G\rho
\end{equation}

\begin{eqnarray}
\label{13}
\dot \phi ^2\left[ \frac 1\phi \frac{d\omega }{d\phi }-\frac
\omega {\phi ^2}-\frac 1G\frac{dG}{d\phi }\frac \omega \phi -\frac
6{G^4}\left( \frac{dG}{d\phi }\right) ^3+\frac 3{G^3}\frac{dG}{d\phi }
\frac{d^2G}{d\phi ^2}\right] +\Box \phi \left[ \frac{2\omega }\phi +\frac
3{G^3}\left( \frac{dG}{d\phi }\right) ^2\right] =   \nonumber  \\
   -\frac 1G\frac{dG}{d\phi }8\pi \rho \left( 4-3\gamma \right)
\end{eqnarray}
\begin{eqnarray}
\label{14}2\frac d{dt}\left( \frac{\dot a}a\right) +3\left( \frac{\dot a}
a\right) ^2+\frac k{a^2}-\left( \frac{\dot a}a\right) \frac 2G
\frac{dG}{d\phi }\dot \phi =-G\,8\pi \rho \left( \gamma -1\right) 
-\frac \omega 2\frac{\dot \phi ^2}\phi G-   \nonumber   \\
   2\left( \frac 1G\frac{dG}{d\phi }\right) ^2\dot \phi
+\frac 1G\frac{d^2G}{d\phi ^2}\dot \phi ^2+\frac 1G\frac{dG}{d\phi }\ddot
\phi
\end{eqnarray}
Note that the solutions of these equations, as remarked by Weimberg \cite
{Weimberg} in the case of Brans Dicke Theory, are defined by four
integration constants. 
It is useful to have the spatial equation in
alternative forms, for instance:

\begin{eqnarray}
\label{16}\dot H+H^2+H\frac 1G\frac{dG}{d\phi }\dot \phi +\frac \omega 3
\frac{\dot \phi ^2}\phi G=
\left[ G\,\frac{8\pi \rho }3\left( (2-3\gamma )\frac \omega \phi -
\frac 3{G^3}\left( \frac{dG}{d\phi }%
\right) ^2\right) +
\frac 12\dot \phi ^2\,\Delta \right] \times \nonumber \\
\frac 1{\left[ \frac{2\omega }\phi +\frac 3{G^3}\left( 
\frac{dG}{d\phi }\right) ^2\right] }
\end{eqnarray}
where we have defined $H$ as usual and

\begin{equation}
\label{17}\Delta =-\frac 3{G^2}\left( \frac{dG}{d\phi }\right) ^2\frac
\omega \phi +\frac 1G\frac{d^2G}{d\phi ^2}\frac {2 \omega}{\phi} -\frac 1G%
\frac{dG }{d\phi }\frac 1\phi \frac{d\omega }{d\phi }+
\frac 1G\frac{dG}{d\phi }%
\frac \omega {\phi ^2}
\end{equation}

The derivation of general barotropic solutions were done only for the case
of generalized BD theories {\it i.e.} $G(\phi )=1/\phi $. The most salient
ones were derived by Nariai \cite{Nariai}, O'Hanloon and Tupper \cite
{O'Hanloon}, Gurevich, Finkelstein and Ruban \cite{Gurevich},
Lorentz-Petzold \cite{Lorentz}, Barrow \cite{Barrow1}, Barrow and Mimoso
\cite{Barrow2} and Mimoso and Wands \cite{MW}. 
Recently, a complete cualitative study of the behavior
of Scalar-Tensor theories was also presented \cite{Alimi}. In what follows we
generalize the method described in \cite{MW} for the equations (9), (10)
and (11). In generalized BD cases it was shown that the change of variables

\begin{equation}
\label{18}X=a^2\phi
\end{equation}

\begin{equation}
\label{19}Y=\int \sqrt{\frac{2\omega (\phi )+3}3}\frac{d\phi }\phi
\end{equation}
together with the introduction of the conformal time defined by the
differential relation:

\begin{equation}
\label{20}dt=a\ d\eta
\end{equation}
allows to rewrite the BD field equations as:

\begin{equation}
\label{21}(X^{\prime })^2+4\ k\ X^2-(Y^{\prime }X)^2=4\ M\ X\ a^{4-3\gamma }
\end{equation}

\begin{equation}
\label{22}(Y^{\prime }X)^{\prime }=M\ (4-3\gamma )\ \sqrt{\frac 3{2\omega +3}%
}\ a^{4-3\gamma }
\end{equation}

\begin{equation}
\label{23}X^{\prime \prime }+4kX=3\left( 2-\gamma \right) M\ a^{4-3\gamma }
\end{equation}
where the density of the barotropic fluid has been written as $\rho =3M\ /\
8\pi a^{3\gamma }$ and the prime denotes differentiation with respect to $%
\eta $. In the general case given by the system (9-10-11) the leading idea is
to retain the simplicity of the transformed system by asking for a suitable
choice of new variables. So we propose them in the form:

\begin{equation}
\label{24}X=\frac{a^2}Gj(\phi )
\end{equation}

\begin{equation}
\label{25}Y=\int \alpha (\phi )\frac{d\phi }\phi
\end{equation}
where $j$ and $\alpha $ ought to be selected in order to maintain the form
of (17-18-19) and have to reduced to their particular values for BD theories
($j=1$ and $\alpha =\sqrt{\frac{2\omega (\phi )+3}3}$) when $G=1/\phi $. \
So, computing all the necessary terms of the transformed system we obtain
two constraint equations (we shall show them in the vacuum case):

\begin{equation}
\label{26}\frac 1\alpha \frac{d\alpha }{d\phi }-\frac 1\phi -\frac 1G\frac{%
dG }{d\phi }+\frac 1j\frac{dj}{d\phi }=\frac{\left[ \frac{d\omega }{d\phi }%
-\frac \omega {\phi ^2}-\frac 1G\frac{dG}{d\phi }\frac \omega \phi -\frac
6{G^4}\left( \frac{dG}{d\phi }\right) ^3+\frac 3{G^3}\frac{dG}{d\phi }\frac{%
d^2G}{d\phi ^2}\right] }{\left[ \frac{2\omega }\phi +\frac 3{G^3}\left(
\frac{dG}{d\phi }\right) ^2\right] }\
\end{equation}

\begin{equation}
\label{27}\dot \phi ^2\left[ \left( \frac 1G\frac{dG}{d\phi }\right)
^2-\left( \frac \alpha \phi \right) ^2+\left( \frac 1j\frac{dj}{d\phi }%
\right) ^2-\frac 2G\frac{dG}{d\phi }\frac 1j\frac{dj}{d\phi }\right] +\dot
\phi \left[ 4\frac{\dot a}a\frac 1j\frac{dj}{d\phi }\right] =-\frac 23\omega
G\frac{\dot \phi ^2 }\phi
\end{equation}
which allow for a solution to be found in the form:

\begin{equation}
\label{28}j=1
\end{equation}

\begin{equation}
\label{29}\alpha =\sqrt{\left( \frac \phi G\right) ^2\left( \frac{dG}{d\phi }%
\right) ^2+\frac 23\omega G\phi }
\end{equation}
So, defining the variables $X$ and $Y$ as in (20-21) and the conformal
time as in (16) the system of field equations simplifies to a form
analogous to the generalized BD cases. As a matter of fact, the
function $\alpha (\phi )$ becomes the same as in (15) for $G(\phi )=1/\phi $%
. In the general Hyperextended Scalar-Tensor formalism it is necessary to
ask for the positivity of the term under the square root in the definition
(\ref{29}). That was also the case in BD theories \cite{MW} where
$\omega$ must be greater than -3/2. The final expression
of the system is then:

\begin{equation}
\label{30}(X^{\prime })^2+4\ k\ X^2-(Y^{\prime }X)^2=4\ M\ X\ \left(
XG\right) ^{\frac{4-3\gamma }2}
\end{equation}

\begin{equation}
\label{31}(Y^{\prime }X)^{\prime }=-M\ (4-3\gamma )\ \frac 1\alpha \left(
XG\right) ^{\frac{4-3\gamma }2}\frac 1G\frac{dG}{d\phi }\phi
\end{equation}

\begin{equation}
\label{32}X^{\prime \prime }+4kX=3\left( 2-\gamma \right) M\ \left(
XG\right) ^{\frac{4-3\gamma }2}
\end{equation}

\section{Cosmological Solutions}

In this section we sketch how to analytically obtain  
cosmological solutions for different perfect fluid universes.
We follow, using the exact reproduction of the form of the 
field equations obtained in the previous section, 
the work of reference \cite{MW}, which may be seen  
for further details.

\subsection{Vacuum Solutions}

Let us first consider the simplest case. In a vacuum model,
the right hand sides
of equations (\ref{30}), (\ref{31}) and (\ref{32}) are equal to zero. Now, 
we use the
fact that the new equations have the same form as the generalized BD
ones.
So, the work made in \cite{MW}, {\it i.e.} the solutions of the system,
is
completely applicable here, except for the different meaning of the
variables. From (\ref{31}) we have $Y^{\prime }X=c$, constant, and so the
solutions for $X$ may be obtained using (\ref{30}). They are
given by equation (3.20) of reference \cite{MW}.
Note $X(\eta )$ is independent of the particular form of $\omega $ and of 
$G$. As $Y^{\prime }X=c$, this implies that

\begin{equation}
\label{34}Y=\int \sqrt{\left( \frac \phi G\right) ^2\left( \frac{dG}{d\phi }%
\right) ^2+\frac 23\omega G\phi }\frac{d\phi }\phi =\int \frac cX\ d\eta
=I(\eta )
\end{equation}
We can compute this integral because of our knowledge of the dependence of $%
X $ over $\eta $. 
So, given the functions $G(\phi )$ and $\omega (\phi )$, we can
compute $Y(\phi )$ and invert it using our knowledge of the right side
of (\ref{34}) to obtain $\phi (\eta )$. 
Together with $a^2=XG$, this yields the
solution of the problem.

Even without solving these equations for particular values of $%
G(\phi )$ and $\omega (\phi )$ it is possible to obtain some general
conclusions about the nature of the singularity in these vacuum models. When
$X\rightarrow 0$ and $\left( \frac{X^{\prime }}X\right) ^2\rightarrow \infty
$, it can be seen that $\frac{X^{\prime }}X\rightarrow \pm Y^{\prime }$.
Using the definition of the variables it is easy to show that:

\begin{equation}
\label{36}\dot a\rightarrow \frac 12\left[ 1\ \mp \frac 1\alpha \frac \phi G
\frac{dG}{d\phi }\right] \frac{X^{\prime }}X
\end{equation}
and the initial singularity, which is produced when $\dot a\rightarrow \pm
\infty $ can only be avoided in these cases when $\omega \rightarrow 0$ or $%
\left( \frac{dG}{d\phi }\right) ^2\gg \frac{2\omega }3\frac{G^3}\phi $. Note
that in the generalized BD cases only the first condition is obtained \cite
{MW}.

\subsection{Non-Vacuum Solutions: Radiation}

With $\gamma =4/3$ the equation of state becomes that of a radiation fluid.
The two first field equations read in this case as:

\begin{equation}
\label{37}(X^{\prime })^2+4\ k\ X^2-(Y^{\prime }X)^2=4\ M\ X\
\end{equation}

\begin{equation}
\label{38}(Y^{\prime }X)^{\prime }=0
\end{equation}
Note that the second equation retains its form from the vacuum case and this
implies again that $Y^{\prime }X=c$. Using this in (\ref{37}) it is 
possible to
integrate for the variable $X$ and then obtain as above the function 
$I(\eta)$.
Once again, due to the exact reproduction of the form of the equations, 
we
have the same solutions as in the BD case but in the new variables,
equation (3.70) of reference \cite{MW}.
It can be seen in this case that at early times all solutions approach the
vacuum ones. 
Thus, defining $G(\phi )$ and $%
\omega (\phi )$ we can follow again the same logical steps to obtain $a^2$
and $\phi $ as functions of $\eta $.

\subsection{Non-Vacuum Solutions: Stiff Matter Fluid}

Let us finally consider case in which $\gamma =2$. That
election represents a barotropic equation of state given by $p=\rho $. The
field equation become in this case:

\begin{equation}
\label{41}(X^{\prime })^2+4\ k\ X^2-(Y^{\prime }X)^2=\frac{4\ M}G\
\end{equation}

\begin{equation}
\label{42}(Y^{\prime }X)^{\prime }=-2M\ \frac 1\alpha \frac 1X\frac 1{G^2}
\frac{dG}{d\phi }\phi
\end{equation}

\begin{equation}
\label{43}X^{\prime \prime }+4kX=0
\end{equation}
The last equation is identical to the corresponding vacuum equations and so $%
X(\eta )$ is given by the same expressions as in the vacuum case. Besides,
we have an useful relation:
\begin{equation}
\label{45}Y^{\prime }X=\pm \sqrt{A-4\frac MG}
\end{equation}
with $A$ a constant of integration.
This requires that:

\begin{equation}
\label{46}\frac A{4M}\geq \frac 1G
\end{equation}
It can be seen that only for $k=-1$ could $A$ be negative. This
means that $G$ is a negative function. 
In this case an extra solution for $X(\eta)$ arise
in addition to the vacuum ones. From (\ref{45}) 
it can be shown that defining:

\begin{equation}
\label{47}Z(\phi )=\int \sqrt{\left( \frac \phi G\right) ^2\left( \frac{dG}{%
d\phi }\right) ^2+\frac 23\omega G\phi }\frac{d\phi }{\phi \sqrt{A-4\frac MG}%
}=\pm \int \frac 1X\ d\eta
\end{equation}
and

\begin{equation}
\label{48}\sqrt{\left( \frac \phi G\right) ^2\left( \frac{dG}{d\phi }\right)
^2+\frac 23\omega G\phi }=\sqrt{\left( \frac \phi {G_{vac}}\right) ^2\left(
\frac{dG_{vac}}{d\phi }\right) ^2+\frac 23\omega _{vac}G_{vac}\phi }\left[
A-4\frac MG\right] \frac 1c
\end{equation}
the vacuum solutions for $\omega _{vac}$ and $G_{vac}$ carry with the $%
\gamma =2$ solutions for $\omega $ and $G$. 
The behavior of the scale factor and of the scalar
field in the stiff matter universe with coupling $\omega $ and {\em %
gravitational constant} $G$ are the same of those of the vacuum universe
with $\omega _{vac}$ and $G_{vac}$. In this general theory and as we have
two generic functions instead of one in the leading lagrangian we can put
all the dependence on $\phi $ in only one vacuum function if convenient.
Then, proceeding as previously done, we can obtain $\phi (\eta )$ and $%
a(\eta )^2$.

\section{Conclusions}

We have shown how to extend the recently presented procedure by Mimoso and
Wands \cite{MW} to obtain the solutions for a generic coupling
simultaneously with a generic dependence of the {\em gravitational constant}
on the field $\phi $, reducing the whole problem to the solution of a single
integral over the field like in \cite{Barrow1,Barrow2,MW}. This can be done
for all curvatures in vacuum, radiation and stiff matter universes.

The particular case in which the leading lagrangian density of the theory is
(3) may be exploited in this general formalism defining $\omega (\phi
)=-\phi \ /\ 2$ for all the $G(\phi )$'s that still retain the positivity of
the expression under the square root in $\alpha $. That case seems
to be clearly important since only for particular choices of $G(\phi )$
an analytical solution is known \cite{Ritis}. 
Examples of the kind of
results that may be obtained in that way, together with other couplings, 
will be presented 
in a forthcoming work.

A crucial point is to note that in this formalism, to equal $\alpha $
[equation (\ref{29})]
correspond equal solution for the field $\phi$. 
This point actually means
that if a solution for a particular $\omega (\phi)$ in a BD like
theory (say $\omega_{BD}$) is known, and we have as result the
$\phi$ and $a^2$ dependences on $\eta$, we can use the
$\phi (\eta)$ as a solution for a class of Hyperextended
Scalar-Tensor theories {\it i.e.} those which have

\begin{equation}
\left( \frac \phi G\right) ^2\left( \frac{dG}{d\phi }%
\right) ^2+\frac 23\omega G\phi=\frac{2\omega_{BD}+3}3
\end{equation}
and obtain
the $a^2$ dependence in each member of the class by using the $X$ definition.
In this way, we could speak of equivalence classes of scalar-tensor 
gravitation, that may, in principle, be formed by an infinite set of members.
Besides, all members of a given class will predict the same results
for all observable quatities that are functions of $\phi$ and $X$.
So, if
we'd were able to prove that for a given set 
$\left( \phi\,,\,X \right)$ or equivalently $\left( \omega\,,\, G\right)$,
a correct behavior in the weak and strong field tests is obtained,
we'd were proving that not only there is not an unique theory of gravity
with equal predictive observational verified power but an infinite set of 
them.

Let us finally comment on the overall feeling 
that one has after the development of the theory concerning how
much it is like generalized BD cases. It can be seen that, for instance, in
the vacuum cases the solutions behave as a whole like in BD theory
with respect to the initial singularity
provided $G(\phi)$ satisfies mildly restrictive conditions.
In the radiation case, 
the solutions behave like in vacuum in exactly the same way as
in BD. And finally, we have also shown that the solutions for a stiff matter
universe are contained in those of vacuum through a convenient choice
of the functions. We believe that the correct way of thinking in these
similarities is
to understand that generalized Brans-Dicke theories stand as
a particular case of the formalism presented in this paper and so, the
cualitative behavior must be expected as similar.

\section{Acknowledgments}

This work was partially supported by CONICET and UNLP. 
D.F.T. wishes to thank A. Helmi for profitable conversations 
in the development of
the research and to J. P. Mimoso for sending to the author his Ph.D. Thesis.


\begin{thebibliography}{99}
\bibitem{BransDicke}  C. Brans and R.H. Dicke, Phys. Rev. {\bf 124}, 925
(1961)

\bibitem{STG}  P.G. Bergmann, Int. J. Theor. Phys. {\bf 1}, 25 (1968); K.
Nortvedt, Astrophys. J. {\bf 161}, 1059 (1970); R.V. Wagoner, Phys. Rev. D.
{\bf 1},3209 (1970)

\bibitem{Nucleo}  A. Serna, R. Dominguez-Tenreiro and G. Yepes, Astrophys.
J. {\bf 391}, 433 (1992); J.A. Casas, J. Garc\'{\i}a-Bellido and M. Quir\'os,
Phys. Lett. B {\bf 94} (1992); D.F. Torres, Phys. Lett. B {\bf 359},
249 (1995); A. Serna and J.M. Alimi, Phys. Rev. D {\bf 53}, 3087 (1996)

\bibitem{Will}  C. Will, {\it Theory and Experiment in Gravitational Physics}
(Cambridge University Press, Cambridge, England, 1981)

\bibitem{Ext}  D. La and P.J. Steinhardt, Phys. Rev. Lett. {\bf 62}, 376
(1989)

\bibitem{Hyperext}  P.J. Steinhardt and F.S. Ascetta, Phys. Rev. Lett. {\bf %
64}, 2470 (1990)

\bibitem{BHole}  J.D. Barrow, Phys. Rev. D {\bf 46}, 3227 (1992)

\bibitem{MAch}  G.J. Withrow and D.G. Randall, MNRAS {\bf 111}, 495 (1951);
R.H. Dicke Phys. Rev. {\bf 125}, 2163 (1962);
A. Beesham, Gen. Rel. and Grav. {\bf 27}, 15 (1995);
J. Barbour and H. Pfister {\em eds.} {\it Machs Principle: From Newton's 
Bucket to Quantum Gravity} (Birkhauser, 1995)
A. Helmi and D.F.Torres,
Universidad Nacional de La Plata and Leiden University Preprint (1996).

\bibitem{Fradkin}  E. S. Fradkin and A.A. Tseytlin, Nuc. Phys. B {\bf 261},1
(1985)

\bibitem{Barrow1}  J.D. Barrow, Phys. Rev. D {\bf 47}, 5329 (1993)

\bibitem{Barrow2}  J.D. Barrow and J.P. Mimoso, Phys. Rev. D {\bf 50}, 3746
(1994)

\bibitem{MW}  J.P. Mimoso and D. Wands, Phys. Rev. D {\bf 51}, 477 (1995)

\bibitem{LiddleWands}  A.R. Liddle and D. Wands, Phys. Rev. D {\bf 45}, 2665
(1992)

\bibitem{Barrow3}  J.D. Barrow and K. Maeda, Nuc. Phys. B {\bf 341}, 294
(1992)

\bibitem{Diploma}  The study of the general lagrangian (\ref{1}) in relation
with Inflation and formation of bubbles was carried out among others by
N. Sakai and K. Maeda, Prog. of Theor. Phys. {\bf 90}, 1001 (1993). The 
notation in this work follows                                
D.F. Torres, Ms. Thesis, Universidad Nacional de La Plata(1995), 
Unpublished. 

\bibitem{Weimberg}  S.W. Weimberg, {\it Gravitation and Cosmology (John
Wiley and Sons, New York, 1972)}

\bibitem{Nariai}  H.Nariai, Prog. Threor. Phys. {\bf 40}, 49 (1968); {\it %
ibid.} {\bf 42}, 544 (1968)

\bibitem{O'Hanloon}  J. O'Hanloon and B.O.J. Tupper, Nuovo Cimento {\bf 7},
305 (1972)

\bibitem{Gurevich}  L.E. Gurevich, A.M. Finkelstein and V.A. Ruban,
Astrophys. Space Sci. {\bf 22}, 231 (1973)

\bibitem{Lorentz}  D. Lorentz-Petzold, Astrophys. Space Sci. {\bf 98}, 101
(1984); {\it ibid}.{\bf \ 98}, 249 (1984); {\it ibid}. {\bf 106}, 409
(1984); {\it ibid}. {\bf 106}, 419 (1984)

\bibitem{Alimi}  A. Serna and M. Alimi, Phys. Rev. D {\bf 53}, 3074 (1996)

\bibitem{Ritis}  S. Capozziello and R. de Ritis, Phys. Lett. A {\bf 195}, 48
(1994). See also: {\it ibid}. {\bf 177}, 1 (1993) and references therein.

\end{thebibliography}
\end{document}